\documentclass[12pt]{iopart}
\usepackage{graphicx}
\usepackage{xcolor}
\usepackage{bm}
\usepackage{cite}
\usepackage{hyperref}

\newcommand{\Hop}{\hat{H}}
\newcommand{\aop}{\hat{a}}
\newcommand{\aopd}{\hat{a}^\dag}
\newcommand{\nop}{\hat{n}}

\begin{document}

%\jpb

\title[Mobile impurities interacting with a few one-dimensional lattice bosons]{Mobile impurities interacting with a few one-dimensional lattice bosons}

\author{Vasil R. Yordanov and Felipe Isaule$^*$}

\address{School of Physics and Astronomy, University of Glasgow, Glasgow G12 8QQ, United Kingdom}
\ead{felipe.isaulerodriguez@glasgow.ac.uk}

\vspace{10pt}

\begin{abstract}

We report a comprehensive study of the ground-state properties of one and two bosonic impurities immersed in small one-dimensional optical lattices loaded with a few interacting bosons. We model the system with a two-component Bose-Hubbard model and solve the problem numerically by means of the exact diagonalization (ED) method. We report binding energies of one and two impurities across the superfluid (SF) to Mott-insulator (MI) transition and confirm the formation of two-body bound states of impurities induced by repulsive interactions. In particular, we found that an insulator bath induces tightly bound di-impurity dimers, whereas a superfluid bath induces shallower bound states.

\end{abstract}

%\maketitle
%\ioptwocol

\section{Introduction}

The study of impurities in quantum mediums has a long history, dating back to the introduction of the Landau-Pekar polaron~\cite{landau_effective_1948}.
Their study is of relevance in a wide variety of problems, ranging from high-$T_c$ superconductivity~\cite{lee_doping_2006} to impurities in nuclear matter~\cite{kutschera_proton_1993}. More recently, the progress in producing ultracold atomic mixtures~\cite{modugno_two_2002,hadzibabic_two-species_2002,silber_quantum-degenerate_2005,ferrier-barbut_mixture_2014} has generated a renewed interest in impurity physics. The high level of control offered by ultracold atom experiments~\cite{bloch_many-body_2008,chin_feshbach_2010} allows the study of impurities in a great variety of configurations, including systems with bosonic or fermionic statistics, a wide range of interaction strengths, and different dimensions.

The problem of a single impurity immersed in an interacting Bose gas, i.e. the Bose polaron, has received particular attention. Bose polarons have been achieved experimentally in the past decade in both one~\cite{catani_quantum_2012} and three~\cite{jorgensen_observation_2016,hu_bose_2016,pena_ardila_analyzing_2019,yan_bose_2020,skou_non-equilibrium_2021} dimensions, motivating extensive theoretical studies~\cite{rath_field-theoretical_2013,pena_ardila_impurity_2015,shchadilova_quantum_2016,volosniev_analytical_2017,grusdt_bose_2017,parisi_quantum_2017,yoshida_universality_2018,camacho-guardian_landau_2018,pastukhov_polaron_2018-1,ichmoukhamedov_feynman_2019,mistakidis_effective_2019,drescher_real-space_2019,panochko_mean-field_2019,hryhorchak_mean-field_2020,isaule_renormalization-group_2021,brauneis_impurities_2021,gamayun_zero_2020,will_polaron_2021}. However, a consistent theoretical description of Bose polarons is not yet achieved. In particular, the bosonic nature of the medium means that multi-body correlations play a significant role in the regime of strong boson-impurity interactions. The effect of these correlations can not be easily taken into account within perturbative approaches, motivating the use of more sophisticated techniques, such as Quantum Monte Carlo~\cite{pena_ardila_monte_2022}. %Furthermore, the superfluid bath induces a polaron-to-molecule crossover~\cite{rath_field-theoretical_2013} which could even support Efimov cycles in some configurations~\cite{levinsen_impurity_2015,sun_visualizing_2017}. 
In contrast, the classic problem of an impurity immersed in a one-component Fermi gas, i.e. the Fermi polaron, only allows two-body correlations, easing its description. Therefore, while the most prominent features of Bose polarons are now well understood, their behavior at finite temperatures and low dimensions, the onset of bound states, among others, are not yet well described.

Closely related to Bose polarons is the problem of two impurities immersed in a Bose gas. This setup has started to attract increasing theoretical attention due to the richer physics induced by the additional impurity~\cite{casteels_bipolarons_2013,dehkharghani_coalescence_2018,camacho-guardian_bipolarons_2018,naidon_two_2018,mistakidis_induced_2020,mistakidis_many-body_2020,theel_counterflow_2022,brauneis_impurities_2021,will_polaron_2021,panochko_two-_2021,jager_effect_2022,petkovic_mediated_2022,pena_ardila_ultra-dilute_2022}. Indeed, it has been shown that two impurities can form bound quasiparticles usually referred to as bipolarons, which are thought to have a closer connection with unconventional pairing and high-$T_c$ superconductors~\cite{alexandrov_bipolarons_1994,alexandrov_unconventional_2008}. The bath induces attractive effective interactions between impurities, leading to the formation of bipolarons with an average radius much smaller than the average distance between two free impurities, generally of the order of the inter-particle distance of the bath~\cite{casteels_bipolarons_2013}. Moreover, these bipolarons form even in systems with purely repulsive interactions~\cite{casteels_bipolarons_2013}. Naturally, the study of bipolarons is much less developed than that of single polarons.

While most theoretical studies of impurities in ultracold atom gases have focused in homogeneous configurations, as the ones previously mentioned, atoms can also be trapped in optical lattices~\cite{bloch_ultracold_2005}, providing a closer connection with condensed-matter systems~\cite{bloch_many-body_2008,lewenstein_ultracold_2007,gross_quantum_2017}. In particular,  one-dimensional lattices loaded with single spin impurities and a bosonic bath have already been achieved experimentally~\cite{fukuhara_quantum_2013}, and further developments are expected from the progress in realizing lattice atomic mixtures~\cite{gunter_bose-fermi_2006,catani_degenerate_2008}. One particular feature of impurities loaded in optical lattices is that bosonic baths undergo a transition from a superfluid (SF) to a Mott-insulator (MI) phase~\cite{fisher_boson_1989}. In tight lattices where the optical potential depth is sufficiently large to confine the system to the lowest Bloch band~\cite{bloch_many-body_2008}, this SF to MI transition is naturally described by the Bose-Hubbard model~\cite{freericks_phase_1994,jaksch_cold_1998}. In this direction, Hubbard models can be straightforwardly extended to include impurities~\cite{bruderer_polaron_2007}, enabling the study of impurities across the SF to MI transition~\cite{colussi_lattice_2022}. It is worth pointing out that while optical lattices do not support phonon excitations~\cite{bloch_ultracold_2005}, and thus the traditional picture of polarons is not fulfilled, condensates introduce Bogoliubov phonons~\cite{griessner_dark-state_2006} which induce polaron-like features in superfluid baths~\cite{bruderer_polaron_2007}.

Here we are concerned with systems of impurities interacting with a bath of bosons trapped in a one-dimensional optical lattice. We consider tight lattices that can be described by a Hubbard-like model.  
Hubbard models have been employed to study different properties of impurities immersed in one-dimensional lattice Bose baths, including their dynamics~\cite{bruderer_transport_2008,johnson_impurity_2011,massel_dynamics_2013} and correlations~\cite{dutta_variational_2013}.  In particular, the binding energies of one and two fermionic impurities in one-dimensional lattices have been recently studied with the density-matrix renormalization group (DMRG)~\cite{pasek_induced_2019}, which can be potentially examined experimentally with spectroscopy measurements~\cite{jorgensen_observation_2016}. In addition, and as an alternative to Hubbard models, impurities in small lattices have been studied using Hamiltonians with oscillatory potentials~\cite{keiler_state_2018,keiler_doping_2020}, particularly to study the effect of shallow optical potentials and entanglement properties~\cite{theel_entanglement-assisted_2020,keiler_doping_2020,pyzh_entangling_2021}.

In this work, we study one and two bosonic mobile impurities immersed in small one-dimensional lattices loaded with a bath of a few (five to eleven) interacting bosons. We describe the system with a two-component Bose-Hubbard model and perform exact diagonalizations (ED) of the Hamiltonian~\cite{callaway_small-cluster_1990,lin_exact_1993,zhang_exact_2010,raventos_cold_2017}. Even though ED restricts our calculations to small lattices, it gives us access to a wide range of properties while taking into account the complete effect of correlations. As mentioned, the latter is essential to study impurity physics due to the importance of correlations, particularly in one-dimensional systems where fluctuations are enhanced~\cite{mistakidis_cold_2022}.  We perform a comprehensive study of the ground-state properties of the impurities across a wide range of inter-atomic interactions across the bath's SF to MI transition, particularly the binding energies. Our results complement and bridge the gap between recent few-~\cite{keiler_state_2018,keiler_doping_2020} and many-body~\cite{pasek_induced_2019,colussi_lattice_2022} calculations of Bose lattice polarons. In the case of two impurities, we also examine the onset of di-impurity dimer states by examining their sizes, where we find that insulating baths induce tightly bound dimers.

This work is organized as follows. In section~\ref{sec:model} we present our model and technical considerations. In sections~\ref{sec:polaron} and ~\ref{sec:bipolaron}  we examine the problems of one and two impurities, respectively. We present results for the binding energies, von Neumann entropy of the bath, and correlations. We also examine the sizes of the di-impurity bound states. Finally, in section~\ref{sec:conclusions} we provide conclusions and an outlook for future studies.

\section{Model}
\label{sec:model}

\subsection{Hamiltonian}

We consider a tight optical lattice with $M$ sites and loaded with $N_B$ bath's bosons and $N_I$ bosonic impurities which interact through on-site potentials. In the following, the subscripts $B$ and $I$ will denote the bath's bosons and impurities, respectively. We  model the system in consideration with a two-component Bose-Hubbard Hamiltonian
\begin{equation}
    \Hop = \Hop_\mathrm{hop}+ \Hop_\mathrm{int}\,.
    \label{sec:model;eq:H}
\end{equation}
The hopping part describes the tunneling of atoms to the nearest neighbor sites
\begin{equation}
    \Hop_\mathrm{hop} = -\sum_i\sum_{\sigma=b,I} t_\sigma\left(\aopd_{i,\sigma}\aop_{i+1,\sigma}+\textrm{h.c.}\right)\,,
    \label{sec:model;eq:Hhop}
\end{equation}
where $\aopd_{i,\sigma}$ ($\aop_{i,\sigma})$ creates (annihilates) a particle $\sigma=B,I$ at site $i$ and $t_\sigma>0$ are the tunneling parameters. The interacting part describes the on-site interactions between atoms

\begin{equation}
    \Hop_\mathrm{int} =\frac{U_{BB}}{2}\sum_i \nop_{B,i}\left(\nop_{B,i}-1\right)+U_{BI}\sum_i\nop_{B,i}\nop_{I,i}\,,
    \label{sec:model;eq:Hint}
\end{equation}
where  $\nop_{i,\sigma}=\aopd_{i,\sigma}\aop_{i,\sigma}$ is the number operator and $U_{BB}$ and $U_{BI}$ are the strengths of the boson-boson and boson-impurity interactions, respectively. The boson-boson interaction is repulsive ($U_{BB}>0$), while the boson-impurity interaction can either be repulsive or attractive. However, we focus mostly on systems with $U_{BI}>0$ because they show richer physics. Note that the impurities do not interact among themselves. 

In this work, we consider lattices loaded with one or two impurities ($N_I=1,2$) and a bath with a unit filling ($\nu_B=N_B/M=1$). We consider that all the atoms have equal masses, and thus $t_B=t_I$~\cite{lewenstein_ultracold_2007}.  We consider lattices with five to eleven sites and periodic boundary conditions. The latter enables us to better extrapolate our results to infinite lattices. Nevertheless, we report a few results for non-periodic lattices in ~\ref{app:n_open}. We also stress that due to the few-body nature of our calculations, any significant change of properties will necessarily show a smooth crossover instead of a well-defined transition.

\subsection{Exact diagonalization}

We extract ground-state properties of Hamiltonian~(\ref{sec:model;eq:H}) by means of the ED method~\cite{callaway_small-cluster_1990,lin_exact_1993}. We work with the usual Fock basis where each state corresponds to a specific distribution of the particles in the lattice
\begin{equation}
 |\alpha_B,\alpha_I\rangle = |n_{B,1},...,n_{B,M}\rangle |n_{I,1},...,n_{I,M}\rangle\,,   
    \label{sec:model;eq:state}
\end{equation}
where $n_{\sigma,i}$ is the number of $\sigma$ bosons in site $i$ and $\sum_i^M n_{\sigma,i} = N_\sigma$. We perform the diagonalizations with the well-known \texttt{ARPACK} package~\cite{lehoucq_arpack_1998}. From the diagonalization we extract the ground-state energy $E_{N_I}$ and the coefficients $c_{\alpha_{B}\alpha_{I}}$ of the ground-state wavefunction
\begin{equation}
    |\Psi_{0}\rangle = \sum_{\alpha_B,\alpha_I} c_{\alpha_{B}\alpha_{I}} |\alpha_B,\alpha_I\rangle \,,
    \label{sec:model;eq:Psi0}
\end{equation}
which we employ to compute the physical properties of interest. For detailed reviews on the ED method for Bose-Hubbard models, we refer to Refs.~\cite{zhang_exact_2010,raventos_cold_2017}.

\subsection{Experimental realization}

A many-body experimental analogue of our setup starts by preparing one bosonic atomic species, such as $^{39}$K or $^{87}$Rb, into two different hyperfine states~\footnote{We choose the same atom to simulate equal tunnelings $t_B=t_I$, but configurations with $t_B\neq t_I$ can be realized with different atomic species.}. The atoms need to be trapped in a one-dimensional optical lattice, similar to recent experiments with two-component bosonic mixtures~\cite{cabrera_quantum_2018}. To simulate isolated mobile impurities, the experiment needs to introduce a large population imbalance, which can be achieved by transferring atoms in one hyperfine state to the other with a radio-frequency pulse. This approach has been used to study Bose polarons in homogeneous gases~\cite{jorgensen_observation_2016}.

The tight optical lattices considered in this work are realized by considering optical potentials with depths $V_0$ much larger than the recoil energy $E_R=\hbar^2 k^2/2m$, where $k=2\pi/\lambda$ and $\lambda$ is the laser's wavelength~\cite{bloch_many-body_2008}. Indeed, lattices with $V_0\gg E_R$ are easily achievable by controlling the laser's intensity~\cite{stoferle_transition_2004}. In typical lattices with a laser's wavelength of $\approx$ 500--1000 nm and a ratio of $V/E_R\approx 10$, the parameters examined in this work can be realized with scattering lengths in the range $|a_s|=$10--200$a_0$~\cite{fukuhara_mott_2009}, where $a_0$ is the Bohr's radius.

Finally, we stress that while the specific few-body setup considered here is difficult to achieve in current experiments, there has been important progress in controlling systems with a few atoms, especially in one dimension~\cite{serwane_deterministic_2011,wenz_few_2013}. While this progress has been achieved mostly with fermionic atoms, further developments are expected in the near future, which has motivated studies of the crossover between few- and many-body physics~\cite{zinner_exploring_2016}, as in this work. We also note that trapping techniques can achieve ring geometries~\cite{henderson_experimental_2009}, which can simulate the periodic boundary conditions considered in the main text, and box configurations~\cite{gaunt_bose-einstein_2013} as considered in~\ref{app:n_open}.

\section{One impurity}
\label{sec:polaron}

We first consider the problem of a single impurity ($N_I=1$). In the following, we study the impurity's binding energy, the effects of the impurity on the bath, and correlations between the bath and the impurity.

\subsection{Binding energies}
\label{sec:polaron;sub:Ep}

The impurity's binding energy $E_p$ corresponds to the energy required to add one impurity to a bath~\cite{jorgensen_observation_2016}. It reads
\begin{equation}
    E_p = E_1-(E_0+E_I)\,,
\label{sec:polaron;sub:Ep;eq:Ep}
\end{equation}
where $E_0$ and $E_1$ correspond to the ground-state energies of the system with zero and one impurity, respectively, and $E_I$ is the energy of a single free impurity. We note that even though the rigorous picture of polarons is not fulfilled in optical lattices, particularly with insulating baths, $E_p$ is still also referred to as the polaron energy for the close connection with Bose polarons~\cite{pasek_induced_2019,colussi_lattice_2022}.

We calculate $E_0$ and $E_1$ numerically  with ED for finite interaction strengths by solving the problem with $N_I=0$ and $N_I=1$, respectively.
In contrast, the energy of a free impurity is simply $E_I=-2t_I$ as a result of the dispersion relation of a free particle in a lattice $\epsilon_I(q)=-2t_I\cos(qa)$~\cite{lewenstein_ultracold_2007}, where $q$ is the momentum and $a$ is the distance between sites.

\begin{figure*}[t]
\centering
\includegraphics[scale=0.75]{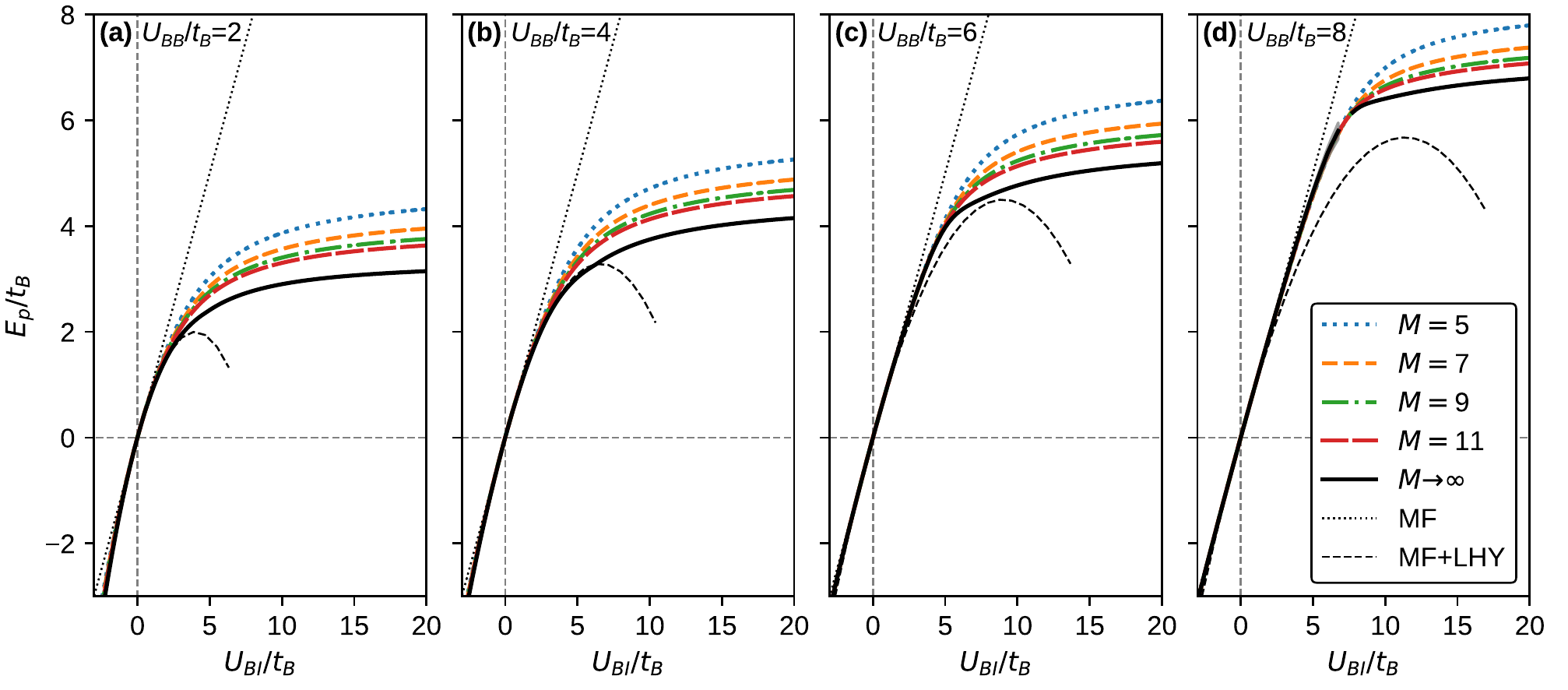}
\caption{Binding energy $E_p$ as a function of $U_{BI}/t_B$ for $\nu_B=1$, $N_I=1$, and $U_{BB}/t_B=2.0 \textrm{ (a)},4.0 \textrm{ (b)},6.0\textrm{ (c)},8.0\textrm{ (d)}$. The colored lines correspond to ED calculations for lattices with five (dotted blue), seven (dashed orange), nine (dash-dotted green), and eleven (long-dashed red) sites. The black solid lines correspond to extrapolations to infinite lattices~(\ref{sec:polaron;sub:Ep;eq:fit}) and the underneath gray regions to the corresponding errors (the segments without lines correspond to regions where the minimization fails).  The thin black dotted and dashed lines correspond to the MF~(\ref{sec:polaron;sub:Ep;eq:Ep_MF}) and MF+LHY solutions~(\ref{sec:polaron;sub:Ep;eq:Ep_pert}--\ref{sec:polaron;sub:Ep;eq:Ep_LHY}), respectively. }
\label{sec:polaron;sub:Ep;fig:Ep_UBI}
\end{figure*}

We show binding energies as a function of the boson-impurity interaction strength $U_{BI}$ for selected bath's parameters $U_{BB}/t_B$ in figure~\ref{sec:polaron;sub:Ep;fig:Ep_UBI}. We note that the SF to MI phase transition of the bath in the thermodynamic limit occurs at $U_{BB}/t_B\approx$ 3.2--3.9 for $\nu_B=1$~\cite{kashurnikov_mott-insulator-superfluid-liquid_1996,kuhner_phases_1998,elstner_dynamics_1999,dos_santos_quantum_2009}. Therefore, we can assume that panel (a) considers a bath in the SF phase, panels (c) and (d) consider baths in the MI phase, and panel (b) considers an intermediate state~\footnote{As mentioned, finite lattices do not show a well-defined phase transition~\cite{raventos_cold_2017}. Instead, finite lattices show a continuous crossover from an SF to a MI phase as $U_{BB}$ increases.}. We also stress again that we focus primarily on repulsive boson-impurity interactions $U_{BI}>0$, as strong attractive interactions simply collapse the few-body system to one site, which cannot be easily connected to a many-body configuration.

We compare our results with the perturbative solution for $t_B=t_I$ and weak coupling~\cite{parisi_quantum_2017,volosniev_analytical_2017,pasek_induced_2019}
\begin{equation}
    E_p = E_p^{\mathrm{(MF)}}+E_p^{\mathrm{(LHY)}}\,,
    \label{sec:polaron;sub:Ep;eq:Ep_pert}
\end{equation}
where
\begin{equation}
     E_p^{\mathrm{(MF)}} =  U_{BI}\nu_B\,,
    \label{sec:polaron;sub:Ep;eq:Ep_MF}
\end{equation}
is the mean-field (MF) solution and~\cite{pasek_induced_2019}
\begin{equation}
    E_p^{\mathrm{(LHY)}} = \frac{U^2_{BI}}{U_{BB}}\left(\frac{1}{2}-\frac{1}{\pi}\arctan\left(\sqrt{\frac{2\,t_B}{U_{BB}n_B}}\right) \right)\,,
    \label{sec:polaron;sub:Ep;eq:Ep_LHY}
\end{equation}
is a Lee-Huang-Yang (LHY)-type correction. We note that the LHY correction is only valid for superfluid baths, and thus we show it in all panels for completeness. Moreover, the MF and LHY solutions are not valid for a collapsing system, as they model a uniform gas.

We also provide estimations for infinite lattices by fitting our results to the function~\cite{raventos_cold_2017}
\begin{equation}
   f(M)=AM^{-b}+C\,, 
    \label{sec:polaron;sub:Ep;eq:fit}
\end{equation}
where $C$ provides the energy for $M\to\infty$. We find the coefficient with a  non-linear least squares algorithm~\cite{branch_subspace_1999}. We stress, however, that for a few interaction strength choices, Eq.~(\ref{sec:polaron;sub:Ep;eq:fit}) does not provide a good approximation for $E_p(M)$ if $M$ is small (generally, if $M \le 8$). Therefore, because we only consider lattices with five to eleven sites, the least-square minimization can fail in some cases.

As already known from other works~\cite{pasek_induced_2019}, we first note that the energy $E_p$ is positive for $U_{BI}>0$, and thus it requires energy to add an impurity to the bath. In contrast, for $U_{BI}<0$ the binding energy is negative, signaling the formation of bound states between the impurity and the bath. 

In all the cases examined, for weak boson-impurity interactions $|U_{BI}/t_B|<3.0$ the results are roughly independent of the lattice's size, showing that $M$ does not play a significant role. Moreover, in this regime, there is a good agreement with the MF solution, as expected. As $U_{BI}>0$ increases, quantum fluctuations become more important and thus the numerical results deviate from the MF solution. The LHY-type correction is able to provide a good description up to $U_{BI}/t_B\approx 5.0$ for $U_{BB}/t_B=2.0,4.0$, but it is ultimately unable to describe systems with stronger interactions. Note that the correction perform even worse than the MF solution for $U_{BB}/t_B=6.0,8.0$. As mentioned, this is expected, as Eq.~(\ref{sec:polaron;sub:Ep;eq:Ep_LHY}) is not suitable for describing Mott baths. On the other hand, in the shown attractive region there is good agreement between all solutions. However, we do not show the collapsing region where the comparisons with the analytical solutions are not valid.

For strong repulsive boson-impurity interactions $U_{BI}/t_B>5.0$ the binding energy shows a clear dependence on the lattice's size, with $E_p$ decreasing as $M$ increases. However, the results show a nice convergence with $M$. Indeed, the overall shape of $E_p$ is already reproduced with five sites. We stress that the obtained behavior of $E_p$ as a function of $U_{BI}$ agrees with DMRG calculations for $\nu_B=2$~\cite{pasek_induced_2019}. We expect that the shown extrapolations for $M\to\infty$ provide good estimations for $E_p$ in large lattices. However, the extrapolations should be compared in the future with robust many-body calculations, such as with Quantum Monte Carlo~\cite{pollet_recent_2012} or DMRG~\cite{schollwock_density-matrix_2005} simulations.

\begin{figure}[t]
\centering
\includegraphics[scale=0.75]{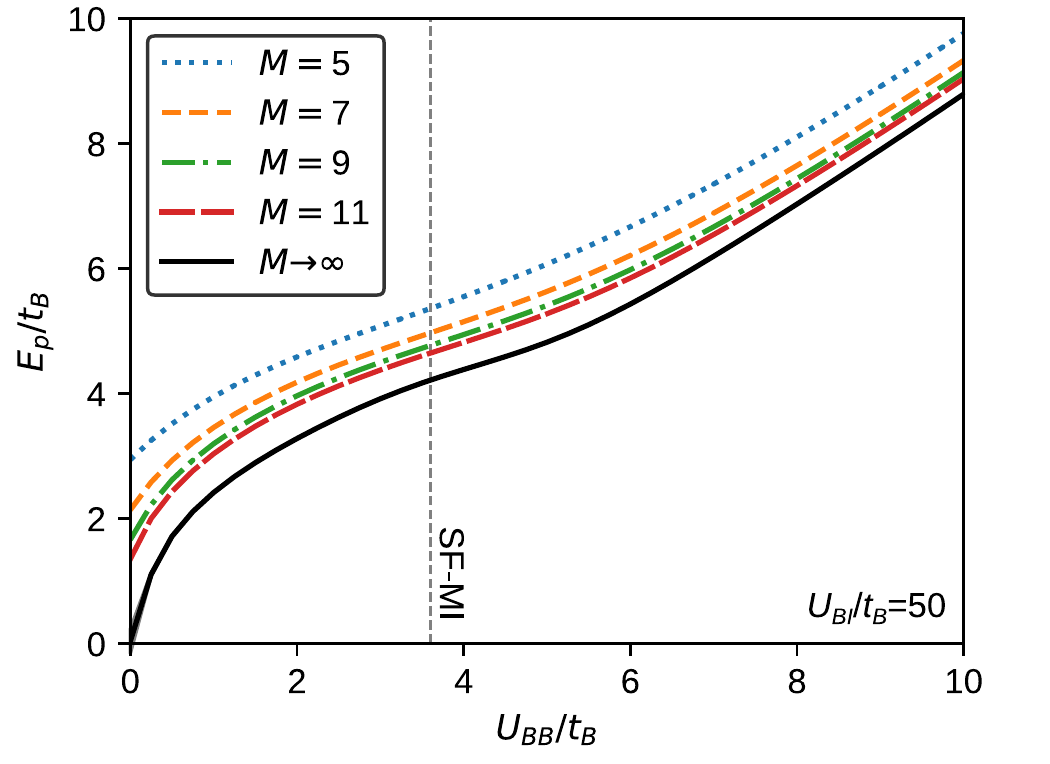}
\caption{Binding energy $E_p$ as a function of $U_{BB}/t_B$ for $\nu_B=1$, $N_I=1$, and $U_{BI}/t_B=50$. The colored lines correspond to ED calculations for lattices with five (dotted blue), seven (dashed orange), nine (dash-dotted green), and eleven (long-dashed red) sites. The black solid lines correspond to extrapolations to infinite lattices~(\ref{sec:polaron;sub:Ep;eq:fit}) and the underneath gray regions to the corresponding errors. The vertical dotted gray line indicates the estimated SF-MI phase transition of the bath $U_{BB}/t_B\approx 3.6$~\cite{kuhner_phases_1998}.}
\label{sec:polaron;sub:Ep;fig:Ep_UBB}
\end{figure}

Finally, to study the dependence of the binding energy on $U_{BB}$ for strong repulsive boson-impurity interactions, in figure~\ref{sec:polaron;sub:Ep;fig:Ep_UBB} we show $E_p$ as a function of $U_{BB}$ for $U_{BI}/t_B=50$. As previously shown, the energy increases with $U_{BB}$. In the MI region ($U_{BB}/t_B<3.6$) the binding energy shows an approximately linear dependence on $U_{BB}$, whereas in the SF region ($U_{BB}/t_B>3.6$) it shows an approximately power-law dependence. We also note that the extrapolation for $M\to\infty$ suggests that the binding energy vanishes in the limit of a non-interacting bath ($U_{BB}=0$), which is expected for an infinite lattice.

\subsection{von Neumann entropy}
\label{sec:polaron;sub:bath}

To study how the impurity affects the bosons, we examine the behavior of the von Neumann entropy of the bath~\cite{raventos_cold_2017}. This is defined as
\begin{equation}
    S_B = -\sum_{\alpha_B} |c^{(B)}_{\alpha_B}|^2\ln |c^{(B)}_{\alpha_B}|^2\,
\label{sec:polaron;sub:bath;eq:SB}
\end{equation}
where $c^{(B)}_{\alpha_B}$ are Fock coefficients obtained by tracing out the impurity states $|\alpha_I\rangle$ [see Eq.~(\ref{sec:model;eq:Psi0})]. $S_B$ measures the clustering of the bath in the Fock space. Indeed, $S_B=0$ in a complete insulator state where the bath is described by only one Fock state $|\Psi^{(B)}_{0,\mathrm{MI}}\rangle=|\nu_B,...,\nu_B\rangle$. In contrast, $S_B$ is large in a superfluid bath where there is a superposition of many Fock states~\cite{raventos_cold_2017}.

\begin{figure}[t]
\centering
\includegraphics[scale=0.8]{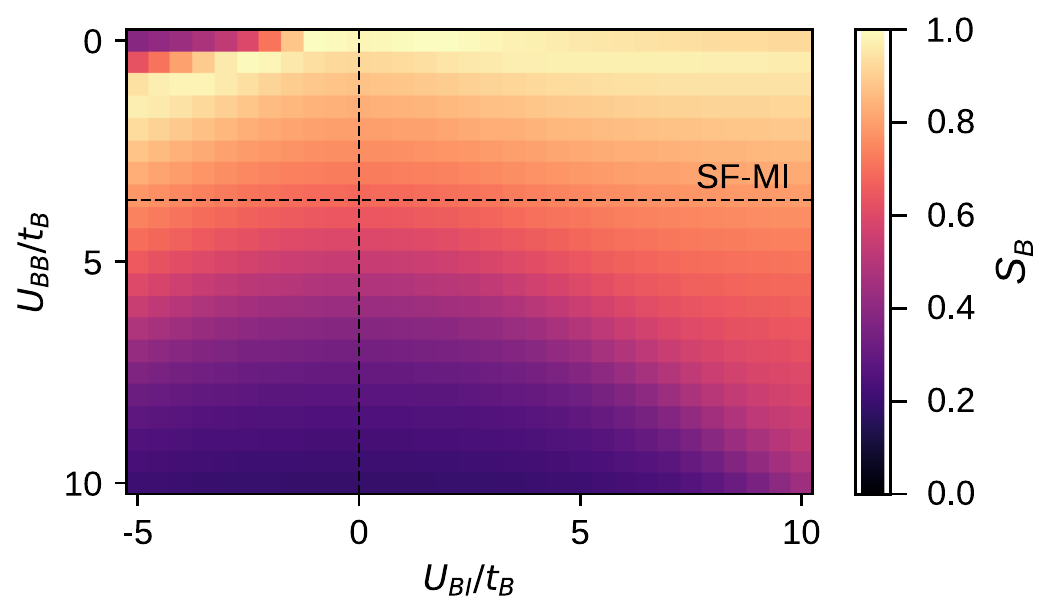}
\caption{Bath's von Neumann entropy $S_{B}$ for $M=9$, $\nu_B=1$, and $N_I=1$ as a function of $U_{BB}/t_B$ and $U_{BI}/t$. The entropy is normalized by its maximum value in the plotted region. The horizontal line indicates the estimated SF-MI phase transition point of the bath $U_{BB}/t_B\approx 3.6$~\cite{kuhner_phases_1998}.}
\label{sec:polaron;sub:bath;fig:SB}
\end{figure}

We show $S_B$ as a function of $U_{BB}$ and $U_{BI}$ in figure~\ref{sec:polaron;sub:bath;fig:SB}.  We consider a lattice with nine sites, but we obtain similar results for other values of $M$. For vanishing boson-impurity interaction $U_{BI}=0$, the bath decouples from the impurity, and therefore $S_{B}$ decreases significantly around the known SF-MI transition point (horizontal line), showing the expected crossover for finite lattices. Indeed, for $U_{BB}/t_B< 3.6$ the entropy takes an approximately constant finite value, signaling a SF phase~\cite{raventos_cold_2017}. In contrast, for $U_{BB}/t_B>3.6$ the entropy slowly vanishes.

$S_B$ shows a clear dependence on the boson-impurity interaction $U_{BI}$, indicating that the bath becomes disturbed by the impurity. As expected, the disruption is larger for stronger boson-impurity interactions. Within the range of interactions examined, by increasing $|U_{BI}|$ the entropy remains large for a wider range of $U_{BB}$ [see region below the SF-MI line]. Our interpretation is that the impurity disturbs the Mott phase by forcing the bosons to move. This delocalizes the bosons in the bath, and thus the bath does not support a complete insulator phase in the whole lattice. This can be expected, as a strongly-repulsive impurity will necessarily prevent the bath to occupy the impurity's site. $S_B$ also vanishes for small $U_{BB}/t_B$ and attractive boson-impurity interactions [see upper-left corner]. This simply corresponds to the collapse of the system to one site, where the bath is described by the few Fock states of the type $|\alpha_B\rangle=|N_B,0,...,0\rangle$.

For further analysis, we examine the condensate fraction of the bath in~\ref{app:Wc}. We also stress that the results of Fig.~\ref{sec:polaron;sub:bath;fig:SB} should not be extrapolated to large lattices where there can be a large distance between bosons and the impurity, particularly the collapsing region.

\subsection{Two-body correlations}
\label{sec:polaron;sub:C2}

To analyze the distribution of the atoms within the lattice, we examine the reduced two-body correlation function~\cite{dutta_variational_2013}
\begin{equation}
C^{(2)}_{i,\sigma'\sigma}=\frac{\langle\psi_{0} |   \aopd_{i, \sigma^{\prime}} \aop_{i, \sigma'} \aopd_{0, \sigma} \aop_{0, \sigma}  | \psi_{0}\rangle}{\langle\psi_{0} | \aopd_{i, \sigma'} \aop_{i, \sigma'}  | \psi_{0}\rangle}\,,
\label{sec:polaron;sub:C2;eq:C2}
\end{equation}
where the denominator is a normalization constant. $C^{(2)}_{\sigma'\sigma}$ measures the average number of bosons of species $\sigma'$ at site $i$ per each boson of species $\sigma$ at a fixed site $i=0$. Note that because the lattice is periodic, the choice of site $i=0$ is arbitrary.

\begin{figure*}[t]
\centering
\includegraphics[scale=0.75]{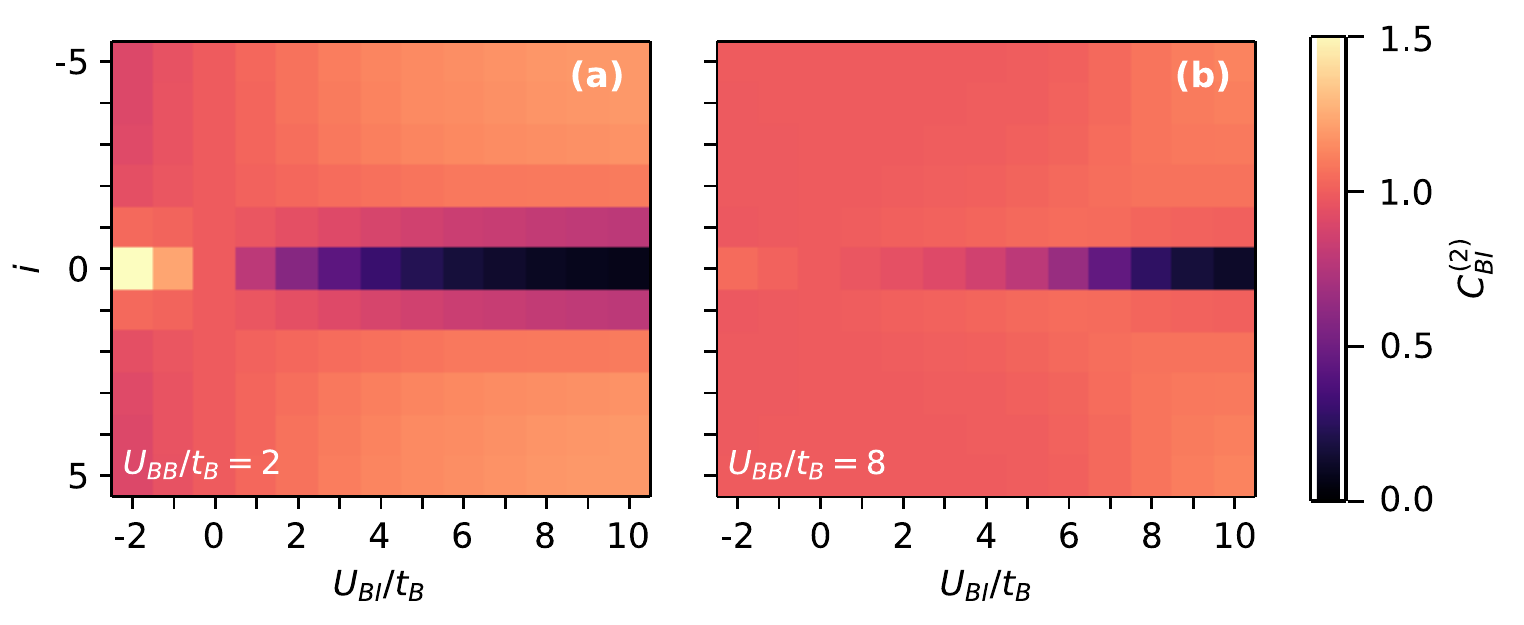}
\caption{Two-body correlations $C^{(2)}_{BI}$ as a function of $U_{BI}/t_B$ and lattice site $i$ for $\nu_B=1$ and $N_I=1$. In panel (a) we consider a superfluid bath with $U_{BB}/t_B = 2.0$ and in panel (b) we consider an insulator bath with $U_{BB}/t_B = 8.0$. We consider lattices with $M=11$.}
\label{sec:polaron;sub:C2;fig:CBI}
\end{figure*}

We show $C^{(2)}_{BI}$ as a function of $U_{BI}$ in figure~\ref{sec:polaron;sub:C2;fig:CBI} for both a weak (a) and strong (b) boson-boson repulsion. The overall behavior of $C^{(2)}_{BI}$ is similar in both panels. For $U_{BI}<0$ the correlations increase around $i=0$ as the impurity attracts the bosons. Indeed, in this region the system shows multi-body bound states, a signature feature of attractive impurities in bosonic baths~\cite{grusdt_bose_2017}. Moreover, for the lattices considered here, we start finding signatures of collapse for $U_{BI}<-U_{BB}$, where the strong boson-impurity interaction tightly binds the rest of the bath to one site. In contrast, as $U_{BI}>0$ increases, the correlations vanish around $i=0$, showing that it is not favorably to have bosons and the impurity at the same site. This \emph{phase separation} between the bath and the impurity is naturally expected for strong boson-impurity repulsion.

The point of separation between the bath and the impurity depends on $U_{BB}$. Indeed, we observe that for $\nu_B=1$ and $t_B=t_I$, the separation occurs approximately at $U_{BI}\approx U_{BB}$. From a mean-field argument, the separation simply occurs when the boson-impurity repulsion surpasses the boson-boson repulsion. However, we again stress that we are not able to locale a well-defined transition due to working with finite small lattices. Therefore, the precise point of phase separation should be studied in the future with more sophisticated many-body approaches.

Finally, it is worth noting that, for large $U_{BI}>0$, while in panel (a) the correlations vanish smoothly around $i=0$, in panel (b) the correlations vanish abruptly. This is because the bath in panel (b) is in an insulator phase. Therefore, it is not favorable for the bath's bosons to move, fixing the impurity to a single site. In contrast, the superfluid bath in panel (a) still allows the impurity to move around $i=0$, even for large boson-impurity repulsion. We further discuss this behavior in~\ref{app:n_open}, where we examine the average occupations of particles in non-periodic lattices.

\section{Two impurities}
\label{sec:bipolaron}

We now turn our attention to the problem of two impurities ($N_I=2$). We perform an analogous study to that of section~\ref{sec:polaron}, with the addition of an examination of the sizes of the induced di-impurity bound states. 

\subsection{Binding energies}
\label{sec:bipolaron;sub:Ebp}

We first examine the binding energy of the two impurities, which is defined as~\cite{camacho-guardian_bipolarons_2018,pasek_induced_2019}
\begin{equation}
    E_{bp} = E_2-2E_1+E_0\,,
\end{equation}
where $E_0$, $E_1$, and $E_2$ are the ground-state energies of the system with zero, one, and two impurities, respectively. We calculate all these energies numerically. In addition, and similarly to the case with one impurity, $E_{bp}$ is also commonly referred to as the bipolaron energy~\cite{pasek_induced_2019}.

\begin{figure*}[t]
\centering
\includegraphics[scale=0.75]{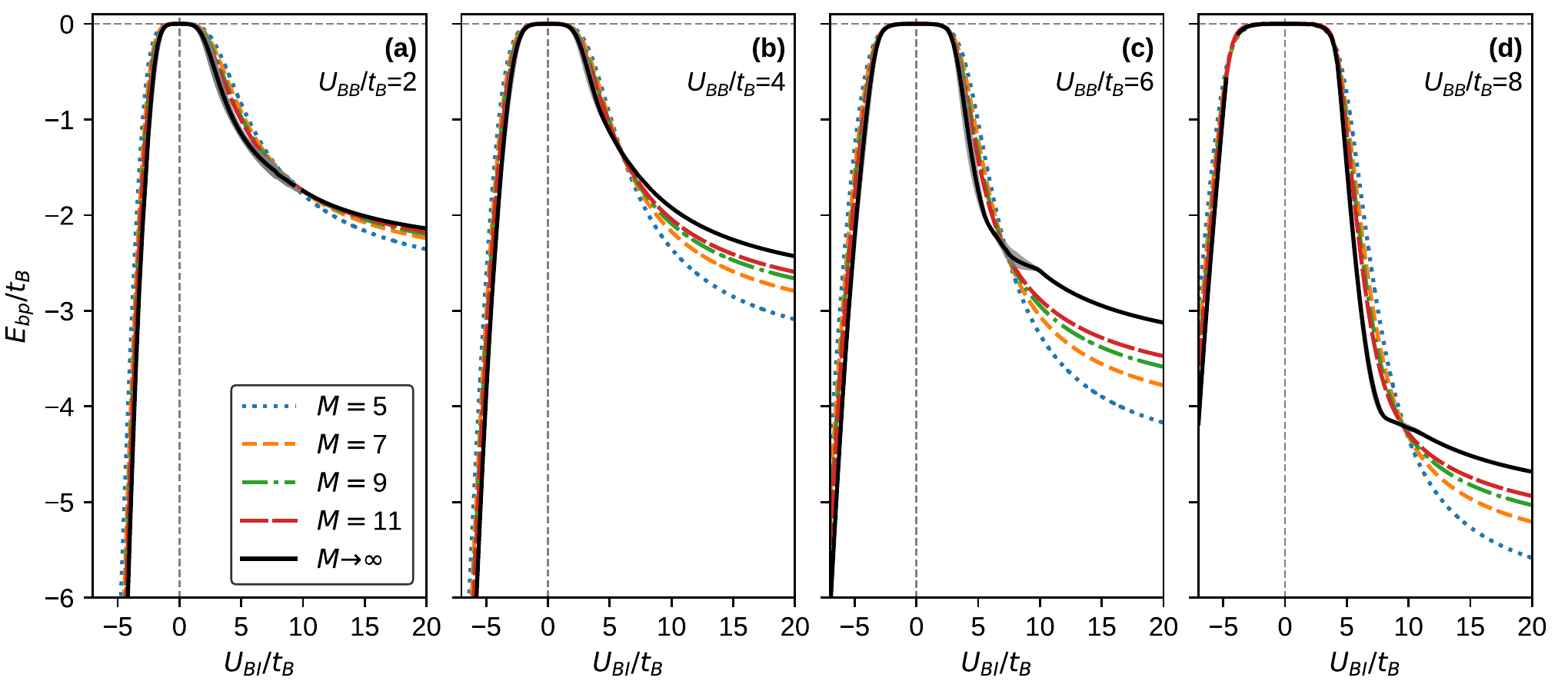}
\caption{Binding energy $E_{bp}$ as a function of $U_{BI}/t_B$ for $\nu_B=1$, $N_I=2$, and $U_{BB}/t_B=2.0 \textrm{ (a)},4.0 \textrm{ (b)},6.0\textrm{ (c)},8.0\textrm{ (d)}$. The colored lines correspond to ED calculations for lattices with five (dotted blue), seven (dashed orange), nine (dash-dotted green), and eleven (long-dashed red) sites. The black solid lines correspond to extrapolations to infinite lattices~(\ref{sec:polaron;sub:Ep;eq:fit}) and the underneath gray regions to the corresponding errors (the segments without lines correspond to regions where the minimization fails).}
\label{sec:bipolaron;sub:Ebp;fig:Ebp_UBI}
\end{figure*}

We show binding energies as a function of the boson-impurity interaction strength for selected bath's parameters in figure~\ref{sec:bipolaron;sub:Ebp;fig:Ebp_UBI}. We show results from ED and extrapolations to $M\to\infty$ obtained from fit~(\ref{sec:polaron;sub:Ep;eq:fit}). In contrast to the one-impurity case, and as also shown by DMRG calculations for $\nu_B=2$~\cite{pasek_induced_2019}, the energy is always negative. This signals the formation of bound states for both attractive and repulsive boson-impurity interactions $U_{BI}$. This is of course expected for attractive interactions. As with one impurity, for attractive $U_{BI}$, the system forms the expected multi-body bound states. In contrast, for repulsive interactions $U_{BI}>0$, two impurities form di-impurity bound states due to the onset of an induced attractive impurity-impurity interaction~\cite{casteels_bipolarons_2013}. We further characterize these di-impurity dimers in the following subsections.

As with one impurity, the binding energy shows a weak dependence on $M$ for weak boson-impurity interactions, while it shows a noticeable dependence on $M$ for large $U_{BI}$. However, the dependence on $M$ changes between weak and strong $U_{BI}>0$. Indeed, in the cases examined $E_{bp}$ decreases with increasing $M$ for small $U_{BI}$, while $E_{bp}$ increases with $M$ for large $U_{BI}$. This means that the di-impurity dimer becomes less bound in large lattices with a strong boson-impurity repulsion. Nonetheless, $E_{bp}$ still shows a nice convergence with $M$, which enables us to extrapolate our results to infinite lattices (black lines).

\begin{figure}[t]
\centering
\includegraphics[scale=0.75]{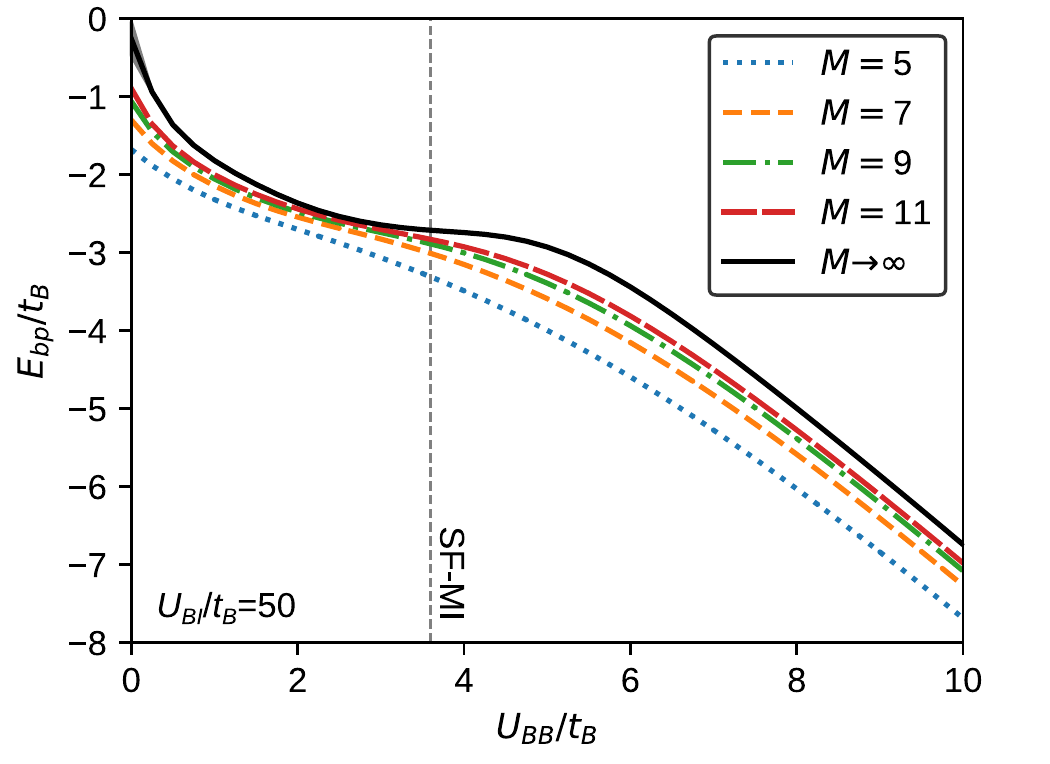}
\caption{Binding energy $E_{bp}$ as a function of $U_{BB}/t_B$ for $\nu_B=1$, $N_I=2$, and $U_{BI}/t_B=50$. The colored lines correspond to ED calculations for lattices with five (dotted blue), seven (dashed orange), nine (dash-dotted green), and eleven (long-dashed red) sites. The black solid lines correspond to extrapolations to infinite lattices~(\ref{sec:polaron;sub:Ep;eq:fit}) and the underneath gray regions to the corresponding errors. The vertical dotted gray line indicates the estimated SF-MI phase transition of the bath $U_{BB}/t_B\approx 3.6$~\cite{kuhner_phases_1998}.}
\label{sec:bipolaron;sub:Ebp;fig:Ebp_UBB}
\end{figure}

We also show the dependence of the binding energy on $U_{BB}$ for large boson-impurity repulsion in figure~\ref{sec:bipolaron;sub:Ebp;fig:Ebp_UBB}. $E_{bp}$ decreases with increasing $U_{BB}$, showing that the di-impurity dimer becomes more bound for larger boson-boson repulsion, as shown previously. In addition, we observe a slight change in the behavior of $E_{bp}$ around the SF to MI transition, with an approximately linear dependence on $U_{BB}$ in the MI phase,  similarly to what we observe with one impurity [see figure~\ref{sec:polaron;sub:Ep;fig:Ep_UBB}]. 

\subsection{Di-impurity sizes}
\label{sec:bipolaron;sub:rbp}

To further examine the formation of bound di-impurity dimers, we study their sizes by calculating the mean average distance between the two impurities~\cite{keiler_doping_2020},
\begin{equation}
    \langle r_{bp}\rangle/a = \langle\Psi_0|\,|i_{I,1}-i_{I,2}|\,|\Psi_0\rangle \,,
\label{sec:bipolaron;sub:rbp;eq:rbp}
\end{equation}
where $a$ is the distance between neighboring sites and $i_{I,N}$ is the position of impurity $N=1,2$. Note that the distance between sites needs to correctly account for the periodic boundary conditions.

\begin{figure*}[t]
\centering
\includegraphics[scale=0.75]{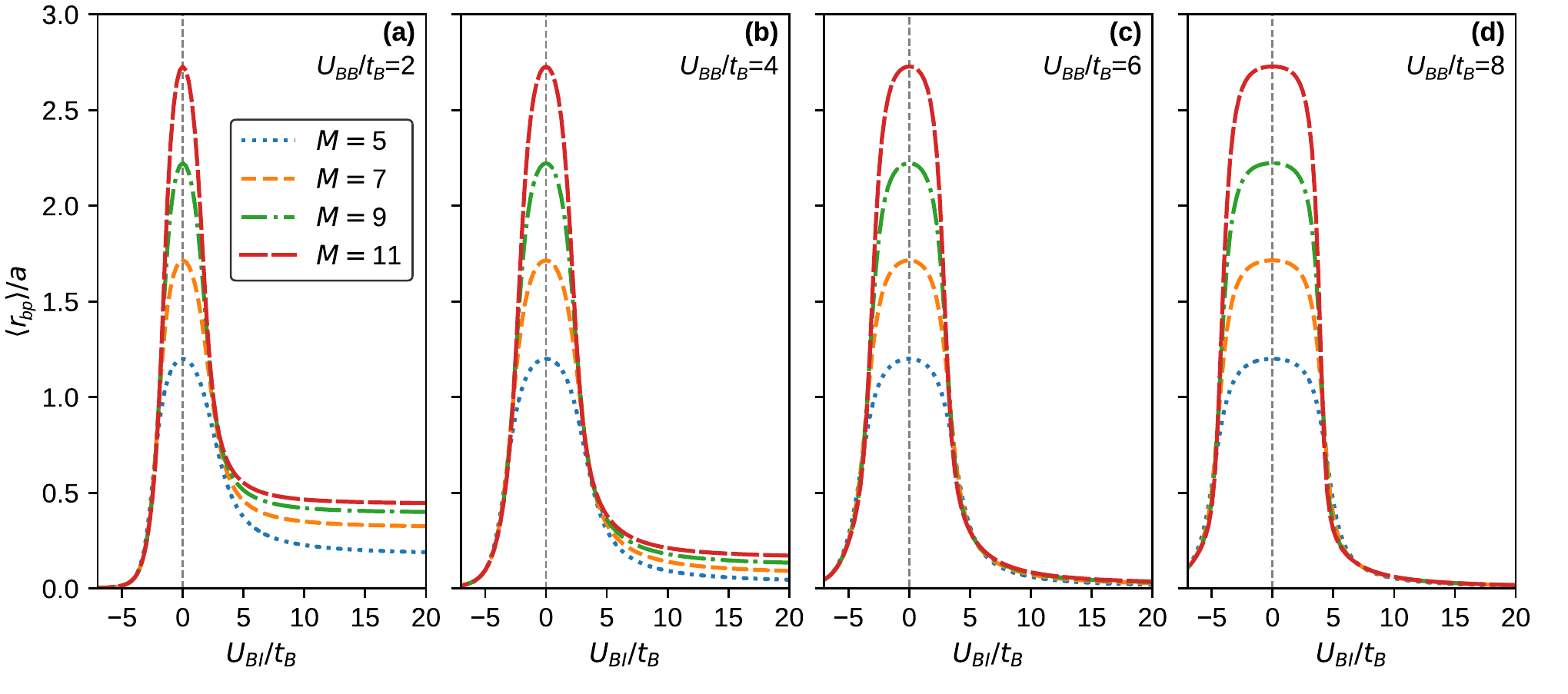}
\caption{Average distance between two impurities $\langle r_{bp}\rangle$ as a function of $U_{BI}/t_B$ for $\nu_B=1$, $N_I=2$, and $U_{BB}/t_B=2.0 \textrm{ (a)},4.0 \textrm{ (b)},6.0\textrm{ (c)},8.0\textrm{ (d)}$. The colored lines correspond to ED calculations for lattices with five (dotted blue), seven (dashed orange), nine (dash-dotted green), and eleven (long-dashed red) sites.}
\label{sec:bipolaron;sub:rbp;fig:rbp_UBI}
\end{figure*}

We show the distance between impurities $\langle r_{bp}\rangle$ as a function of $U_{BI}$ in figure~\ref{sec:bipolaron;sub:rbp;fig:rbp_UBI}. The distance reaches its maximum value for vanishing boson-impurity interaction $U_{BI}=0$, while it decreases with increasing $|U_{BI}|$. The distance at $U_{BI}=0$ simply corresponds to the average separation between two free particles in a lattice with $M$ sites. Therefore, $\langle r_{bp}\rangle$ depends strongly on $M$ around $U_{BI}\approx 0$.  In contrast, $\langle r_{bp}\rangle$ shows a clear convergence with increasing $M$ for large $|U_{BI}|$. 

The distance between impurities rapidly vanishes for $U_{BI}<0$ in all the cases examined, consistent with the collapse of the particles to one site and the formation of a single multi-body bound state. On the other hand, $\langle r_{bp}\rangle$ converges to finite values for large $U_{BI}>0$. Nonetheless, $\langle r_{bp}\rangle$ converges to values smaller than the lattice spacing $a$, signaling the formation of bound states between the two impurities. As we further examine in subsection~\ref{sec:bipolaron;sub:C3}, for $U_{BI}>0$ the impurities do not bind with the bath, and thus this corresponds only to a two-body dimer state between the two impurities. Furthermore, $\langle r_{bp}\rangle$ converges to smaller values as $U_{BB}/t_B$ increases, suggesting that a stronger boson-boson repulsion induces a stronger effective attraction between impurities. Therefore, the impurities become more bound for increasing $U_{BB}$. This is consistent with the increasing of $|E_{bp}|$ with $U_{BB}>0$ reported in figure~\ref{sec:bipolaron;sub:Ebp;fig:Ebp_UBI}.  We also note that similar results for $\langle r_{bp}\rangle$ have been reported in Ref.~\cite{keiler_doping_2020} for lattices with five sites.

\begin{figure}[t]
\centering
\includegraphics[scale=0.75]{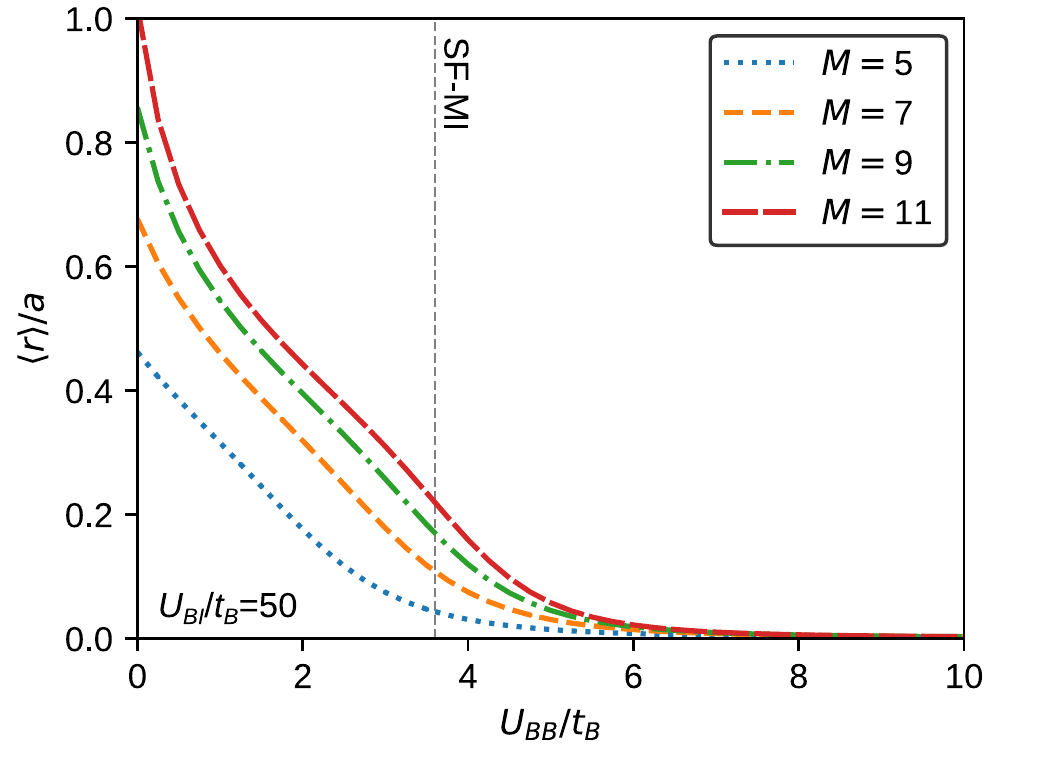}
\caption{Average distance between two impurities $\langle r_{bp}\rangle$ as a function of $U_{BB}/t_B$ for $\nu_B=1$, $N_I=2$, and $U_{BI}/t_B=50$. The colored lines correspond to ED calculations for lattices with five (dotted blue), seven (dashed orange), nine (dash-dotted green), and eleven (long-dashed red) sites. The vertical dotted gray line indicates the estimated SF-MI phase transition of the bath $U_{BB}/t_B\approx 3.6$~\cite{kuhner_phases_1998}.}
\label{sec:bipolaron;sub:rbp;fig:rbp_UBB}
\end{figure}

Finally, in figure~\ref{sec:bipolaron;sub:rbp;fig:rbp_UBB} we show $\langle r_{bp}\rangle$  as a function of $U_{BB}$ for large boson-impurity repulsion. As previously shown, the distance between impurities decreases with increasing $U_{BB}$. Indeed, while in an SF bath there is a finite average distance between the two impurities, in a MI bath this distance vanishes. This means that for large $U_{BB}>0$ the two impurities form tightly bound dimers, with the two impurities localized in the same site. In contrast, for small $U_{BB}>0$ the impurities form shallower bound states. Once again, this is consistent with the behavior of $E_{bp}$ reported in figure~\ref{sec:bipolaron;sub:Ebp;fig:Ebp_UBB}. We further discuss the origin of this behavior in subsection~\ref{sec:bipolaron;sub:C3}.

\subsection{von Neumann entropy}
\label{sec:bipolaron;sub:bath}

\begin{figure}[t]
\centering
\includegraphics[scale=0.75]{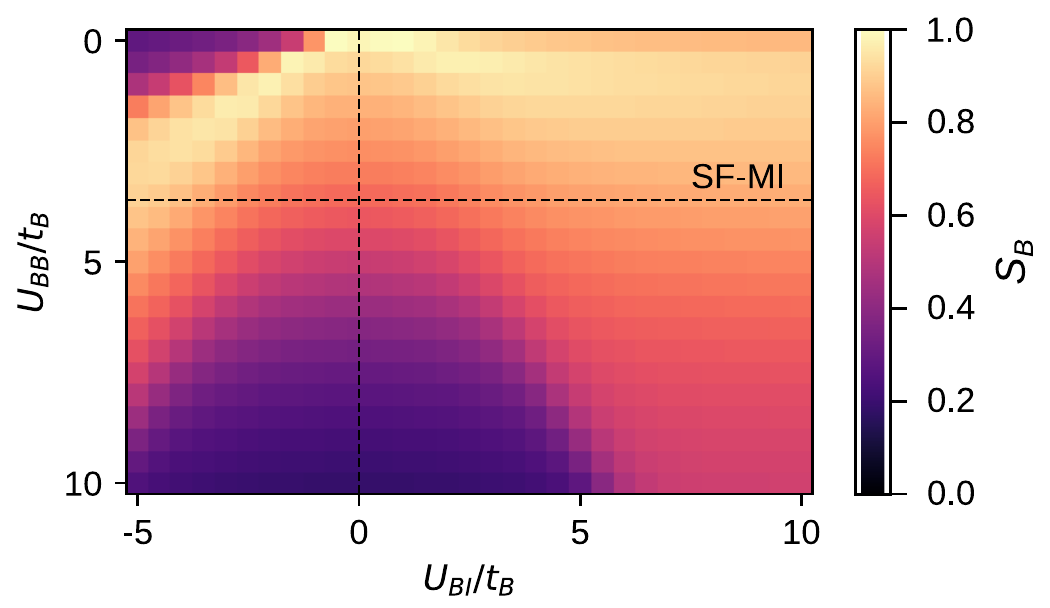}
\caption{Bath's von Neumann entropy $S_{B}$ for $M=9$, $\nu_B=1$, and $N_I=2$ as a function of $U_{BB}/t_B$ and $U_{BI}/t_B$. The entropy is normalized by its maximum value in the plotted region. The horizontal line indicates the estimated SF-MI phase transition point for bath $U_{BB}/t_B\approx 3.6$~\cite{kuhner_phases_1998}.}
\label{sec:bipolaron;sub:bath;fig:Ωc}
\end{figure}

We now examine the impact of the two impurities on the bath. In figure~\ref{sec:bipolaron;sub:bath;fig:Ωc} we show the bath's von Neumann entropy $S_{B}$ [see subsection~\ref{sec:polaron;sub:bath}] in the presence of two impurities. We observe a similar behavior to that induced by only one impurity [see figure~\ref{sec:polaron;sub:bath;fig:SB}]. An increasing $|U_{BI}|$ disturbs the insulator phase, increasing the region with a large $S_B$. Similarly, for strong attractive boson-impurity interaction [top left corner of the figure] the system simply collapses to one site. However, with the additional impurity, there is a greater impact on the bath, as expected.

Within the same range of parameters, figure~\ref{sec:bipolaron;sub:bath;fig:Ωc} shows a noticeably smaller full insulator region than that with one impurity [figure~\ref{sec:polaron;sub:bath;fig:SB}]. Similarly, the collapsing region is larger with two impurities, showing the impact of the additional impurity. This pattern is expected to continue for more impurities, which could be examined in future studies.

As with one impurity, we provide an additional discussion in terms of the condensate fraction in~\ref{app:Wc}.

\subsection{Three-body correlations}
\label{sec:bipolaron;sub:C3}

In the following we examine the reduced correlations, in analogy to the study presented in subsection~\ref{sec:polaron;sub:C2}. However, because we now consider two impurities, in the following we examine the reduced three-body correlation function
\begin{equation}
    C^{(3)}_{ji,\sigma''\sigma'\sigma}=\frac{\langle\psi_{0}|\aopd_{j, \sigma''} \aop_{j, \sigma''}  \aopd_{i, \sigma'} \aop_{i, \sigma'} \aopd_{0, \sigma} \aop_{0, \sigma}  | \psi_{0}\rangle}{\langle\psi_{0} | \aopd_{0, \sigma} \aop_{0, \sigma}  | \psi_{0}\rangle}\,.
\label{sec:bipolaron;sub:C3;eq:C3}
\end{equation}
Similar to the interpretation of $C^{(2)}$ [Eq.~(\ref{sec:polaron;sub:C2;eq:C2})], $C^{(3)}_{\sigma''\sigma'\sigma}$ measures the average number of bosons of species $\sigma''$ at site $j$ and of species $\sigma'$ at sites $i$ for each boson $\sigma$ at site $i=0$. 

\begin{figure}[t]
\centering 
\includegraphics[scale=0.75]{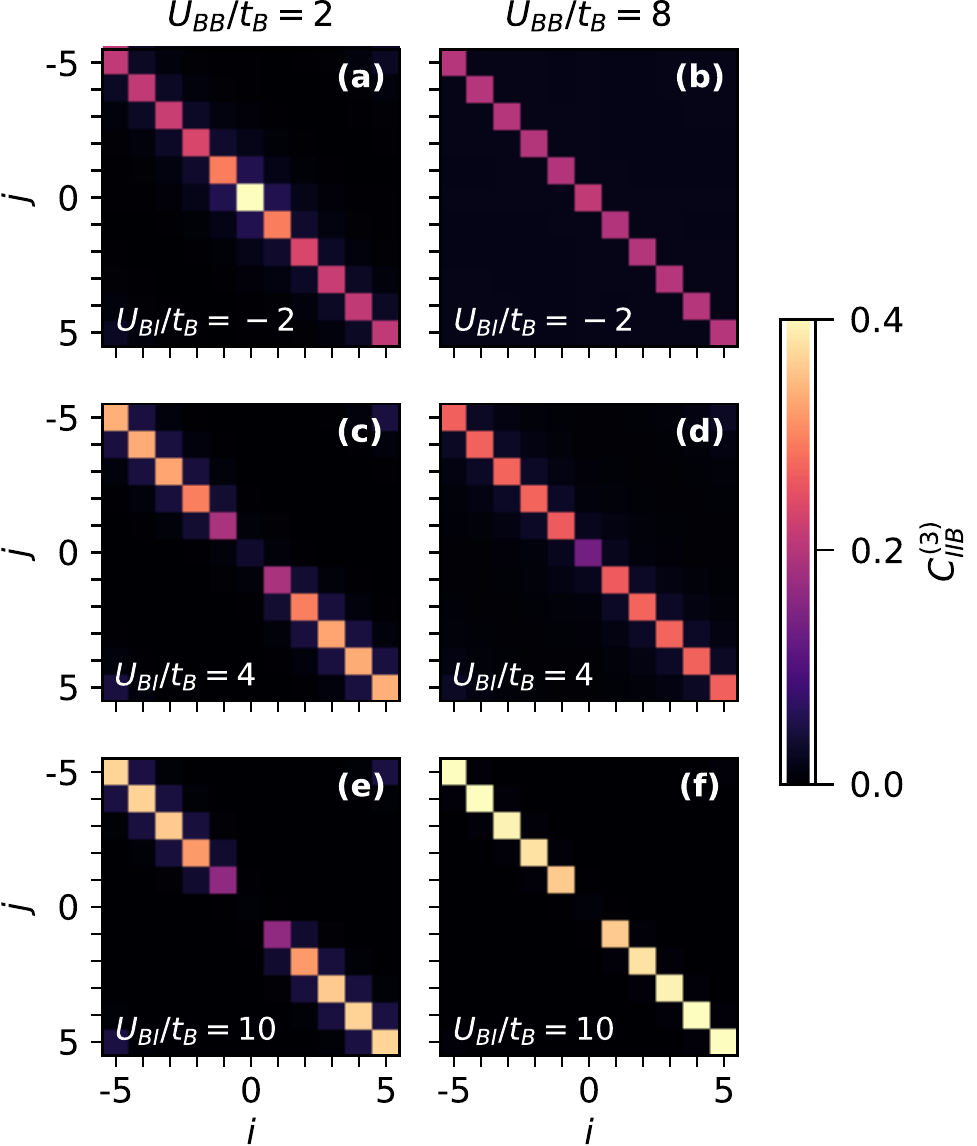}
\caption{Three-body correlation $C^{(3)}_{IIB}$ as a function of the lattice sites $i$ and $j$ for $\nu_B=1$, and $N_I=2$. In panels (a,c,e) we consider a superfluid bath with $U_{BB}/t_B = 2.0$ and in panels (b,d,f) we consider an insulator bath with $U_{BB}/t_B = 8.0$. We consider lattices with $M=11$. We show results for $U_{BI}/t_B=-2.0\textrm{ (a,b), }4.0\textrm{ (c,d), }8.0\textrm{ (e,f)}$.}
\label{sec:bipolaron;sub:C3;fig:CIIB}
\end{figure}

To examine the behavior of the two impurities around the bath's bosons, we show the 
correlations $C^{(3)}_{IIB}$ in figure~\ref{sec:bipolaron;sub:C3;fig:CIIB} for weak (left panels) and strong (right panels) boson-boson interactions. We note that for attractive boson-impurity interactions (top panels), the correlations are larger at $i=j=0$, showing that the bath's bosons attract both impurities to one site. As expected, for strong $U_{BI}<0$, large correlations at $i=j=0$ signal the collapse of the system and the formation of a multi-body bound state. In contrast, for large repulsive boson-impurity interactions (bottom panels), the correlations vanish around $i=j=0$. This means that for $U_{BI}>0$ the impurities are repelled by the bath, forming only a two-body bound state. As with one impurity, the impurities become phase-separated from the bath.

\begin{figure}[t]
\centering
\includegraphics[scale=0.75]{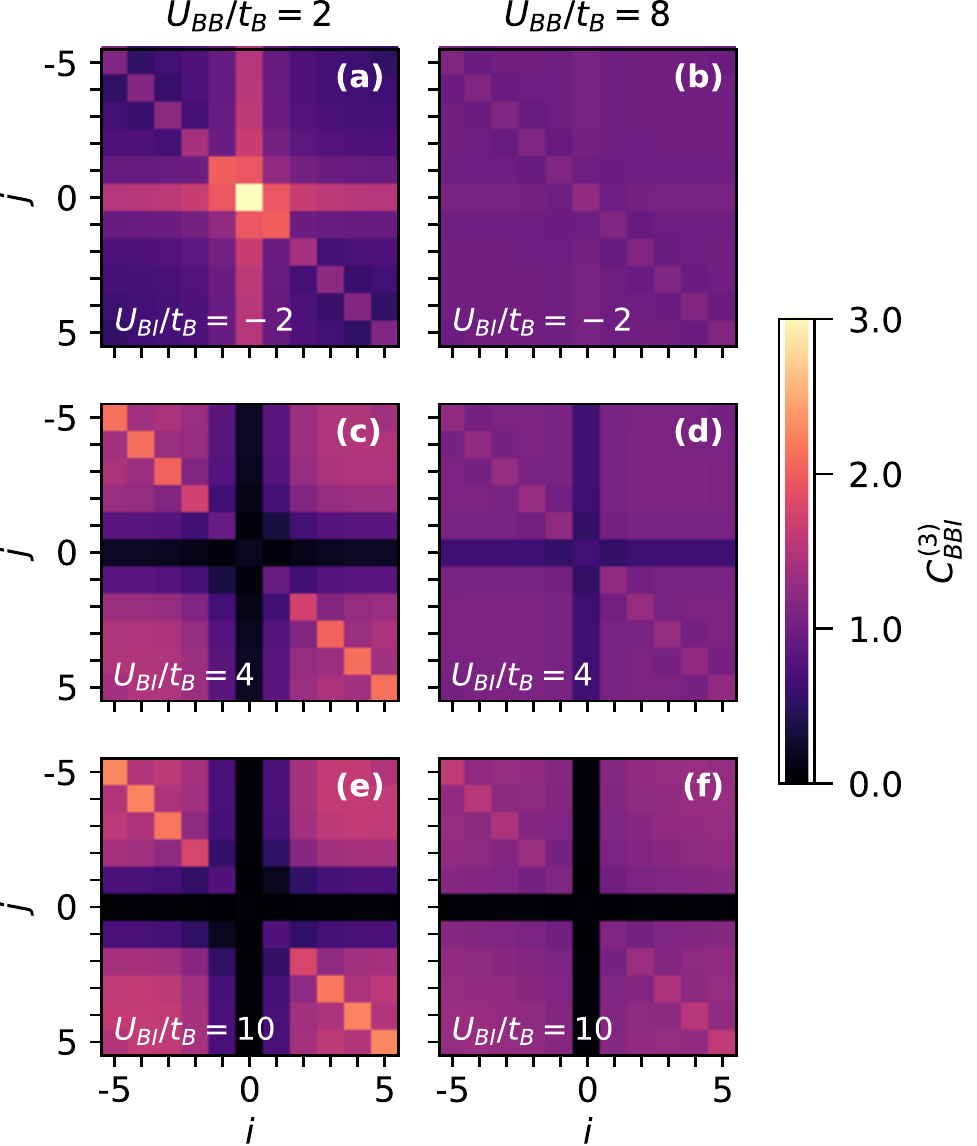}
\caption{Three body correlation $C^{(3)}_{BBI}$ as a function of the lattice sites $i$ and $j$ for $\nu_B=1$, and $N_I=2$. In panels (a,c,e) we consider a superfluid bath with $U_{BB}/t_B = 2.0$ and in panels (b,d,f) we consider an insulator bath with $U_{BB}/t_B = 8.0$. We consider lattices with $M=11$. We show results for $U_{BI}/t_B=-2.0\textrm{ (a,b), }4.0\textrm{ (c,d), }8.0\textrm{ (e,f)}$.}
\label{sec:bipolaron;sub:C3;fig:CBBI}
\end{figure}

To examine the behavior of the bath's bosons around the impurities, we also show $C^{(3)}_{BBI}$ in figure~\ref{sec:bipolaron;sub:C3;fig:CBBI}. We observe that for attractive boson-impurity interactions, the correlations are larger around $i=0$ and $j=0$, signaling again the formation of bound states between the impurities and the bosons. This is similar to what we observe with one impurity in figure~\ref{sec:polaron;sub:C2;fig:CBI} for $U_{BI}<0$. In contrast, as the boson-impurity repulsion increases, the correlations vanish around $i=0$ and $j=0$, showing that the di-impurity dimer is phase-separated from the bath. 
 
Similar to what we observe in figure~\ref{sec:polaron;sub:C2;fig:CBI}, in figure~\ref{sec:bipolaron;sub:C3;fig:CBBI}(e) there is smooth change in $C^{(3)}$ around $i=0$ and $j=0$, whereas in panel (f) there is a sharp transition between vanishing and finite correlations. As expected, in an insulator bath (f) the correlations are roughly constant for $i\neq 0$ and $j\neq 0$. Therefore, we can conclude that, for strong boson-impurity repulsion, an insulator bath induces tightly bound di-impurity dimers because it is not favorable for the bath's bosons to move. In contrast, a superfluid bath enables the impurities to form shallower bound states. We further discuss this behavior in~\ref{app:n_open} where we examine the average occupation of particles in non-periodic lattices.

\section{Conclusions}
\label{sec:conclusions}

In this work, we provided a comprehensive study of ground-state properties of one and two bosonic impurities immersed in a one-dimensional Bose lattice bath. We employed an exact diagonalization method for small lattices, which enabled us to capture the complete effect of fluctuations in the regime of strong interactions. 

We found that our calculations correctly describe the regime of strong interactions, which is not accessible by mean-field calculations. We examined binding energies of one and two impurities across the SF and MI phases of the bath and found that our results are consistent with related works. Similarly, we examined the sizes of the di-impurity bound states in repulsive systems with two impurities and
found that these decrease with increasing boson-impurity repulsion. Indeed, we found that a strongly-repulsive bath induces tightly bound di-impurity dimers, whereas a weakly-repulsive bath induces shallower bound states. We concluded that an insulator bath necessarily induces tightly-bound dimers because it is not favorable for the bath's bosons to tunnel to neighboring sites.

In the future, we intend to study impurities in larger one- and two-dimensional Bose lattice baths by performing Monte Carlo~\cite{pollet_recent_2012} and DMRG~\cite{schollwock_density-matrix_2005} simulations. This will enable us to test our conclusions in many-body scenarios and complement recent perturbative studies~\cite{colussi_lattice_2022}. We also intend to study impurities immersed in lattices loaded with spin $1/2$ fermions and two-component bosons. We expect these more complex baths will induce richer physics, similar to the richer behavior observed by analogous studies in homogeneous gases~\cite{hu_crossover_2022,keiler_polarons_2021,liu_polarons_2022}.

\section*{Acknowledgements}
FI thanks helpful discussions with A. Rojo-Francàs and
B. Juliá-Díaz. FI acknowledges funding from EPSRC
(UK) through Grant No. EP/V048449/1.

\section*{Data availability}
The ED results shown in this article are freely available in Ref.~\cite{yordanov_vasil_r_data_2022}

%\section*{Abbreviations}
%\begin{tabular}{l l}
%    SF & Superfluid\\
%    MI & Mott-insulator\\
%    DMRG & Density-matrix renormalization group\\
%    ED & Exact diagonalization\\
%    MF & mean-field\\
%    LHY & Lee-Huang-Yang
%\end{tabular}

\appendix

\section{Average occupations in open lattices}
\label{app:n_open}

\begin{figure*}[t]
\centering 
\includegraphics[scale=0.75]{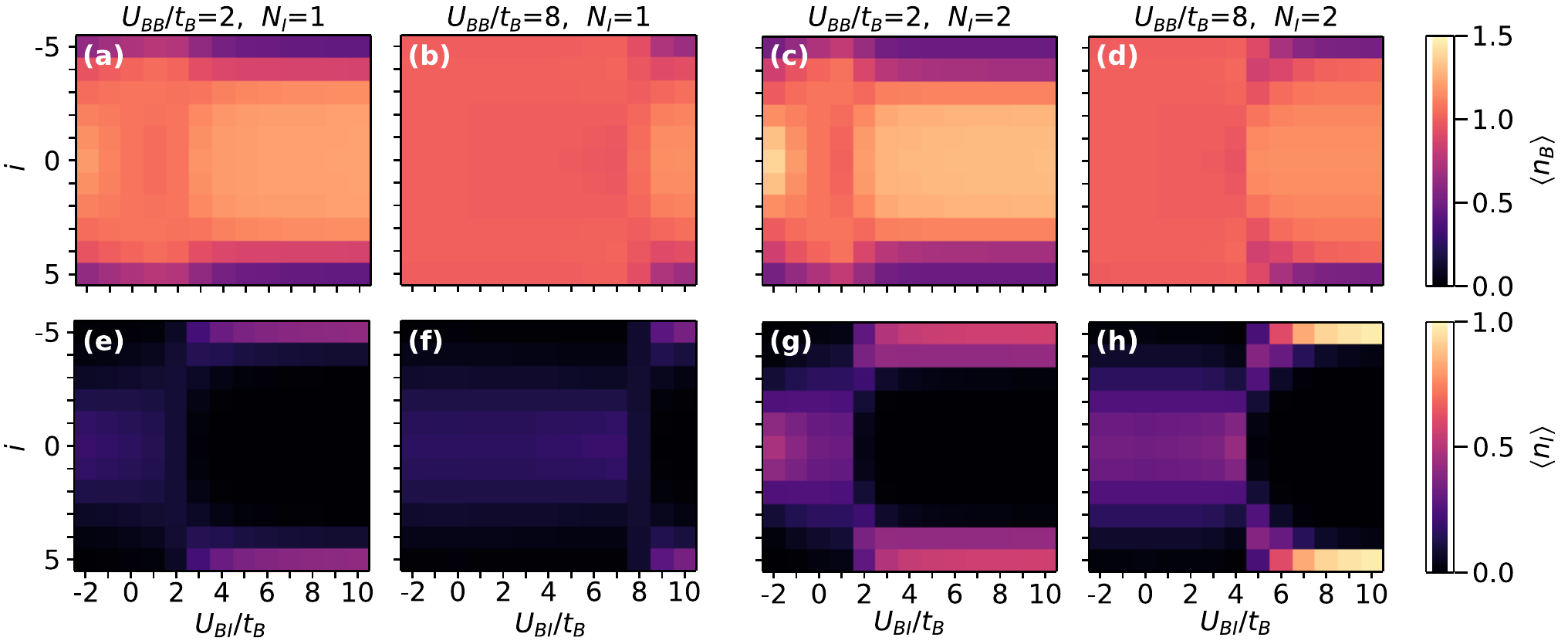}
\caption{Average boson (upper panels) and impurity (bottom panels) occupation per site $\langle n_{i,\sigma} \rangle$ ($\sigma=B,I$) as a function of $U_{BI}/t_B$ and lattice site $i$ for $\nu_B=1$, $N_I = 1 \textrm{ (a,b,e,f), } 2 \textrm{ (c,d,g,h)}$, and $U_{BB}/t_B = 2.0 \textrm{ (a,c,e,g), } 8.0 \textrm{ (b,d,f,h)}$. We consider lattices with $M=11$.}
\label{app:n_open;fig:n}
\end{figure*}

In this appendix, we examine lattices with open (non-periodic) boundary conditions to complement the results of the main text. We study the average occupation of species $\sigma$ per site $i$,
\begin{equation}
    \langle n_{i,\sigma} \rangle = \langle \Psi_0 | \aopd_{i,\sigma}\aop_{i,\sigma} | \Psi_0 \rangle\,.
\label{app:n_open;eq:n}
\end{equation}
We show the average occupations of the bath's bosons and impurities in figure~\ref{app:n_open;fig:n}. We show occupations for lattices with one impurity (left panels) and two impurities (right panels). For weak boson-impurity repulsion $U_{BI}<U_{BB}$, the bath's bosons can occupy the entire lattice, as expected. In contrast, for strong boson-impurity repulsion, the bath's bosons cannot occupy the same site as the impurities, forcing the impurities to move to the boundaries of the lattice.  This behavior has already been reported in shallow open lattices with five sites~\cite{keiler_doping_2020} and in related studies of impurities trapped in one-dimensional harmonically confined Bose gases~\cite{dehkharghani_quantum_2015}. Naturally, this behavior simply corresponds to the phase separation discussed in subsections~\ref{sec:polaron;sub:C2} and~~\ref{sec:bipolaron;sub:C3}. Indeed, we find that the approximate point of phase separation is consistent with that observed for periodic lattices.

In the case of two impurities, one relevant difference between the weakly- and strongly-repulsive bath is that for $U_{BB}/t_B=2$ the two impurities can occupy the two sites at the boundary of the lattice [see the right region in (g)], while for $U_{BB}/t_B=8$ the two impurities occupy a single site [see the right region in panel (h)]. This is consistent with the behavior observed in periodic lattices [see figure~\ref{sec:bipolaron;sub:rbp;fig:rbp_UBB}], where an insulating bath induces tightly bound di-impurity states, while a superfluid bath induces shallower bound states. We also note that analogous occupation profiles have been reported with small open lattices in Ref.~\cite{keiler_doping_2020}.

\section{Condensate fraction}
\label{app:Wc}

To complement the results shown in~\ref{sec:polaron;sub:bath} and~\ref{sec:bipolaron;sub:bath}, in the following we examine the impact of the impurities on the condensate fraction of the bath $\Omega_{c,B}=\lambda_{M}/N_B$. Here $\lambda_{M}$ is the largest eigenvalue of the bath's one-body density matrix~\cite{zhang_exact_2010}
\begin{equation}
    \rho^{(B)}_{ij}=\langle \Psi_{0} |\aopd_{i,B}\aop_{j,B} | \Psi_{0}\rangle \,,
\end{equation}
where $|\Psi_{0}\rangle$ corresponds to a solution given by Eq.~(\ref{sec:model;eq:Psi0}). Even though condensation and superfluidity are related but different phenomena, $\Omega_c$ can be used to identify the phase of a Bose-Hubbard model within ED calculations.~\cite{zhang_exact_2010}. Indeed, in the MI phase $\Omega_{c,B}\approx 0$ as particles necessarily occupy different Bloch states. In contrast, in the SF phase particles can populate the lowest state and thus $\Omega_{c,B}\approx 1$~\cite{roth_phase_2003}.

\begin{figure*}[t]
\centering
\includegraphics[scale=0.75]{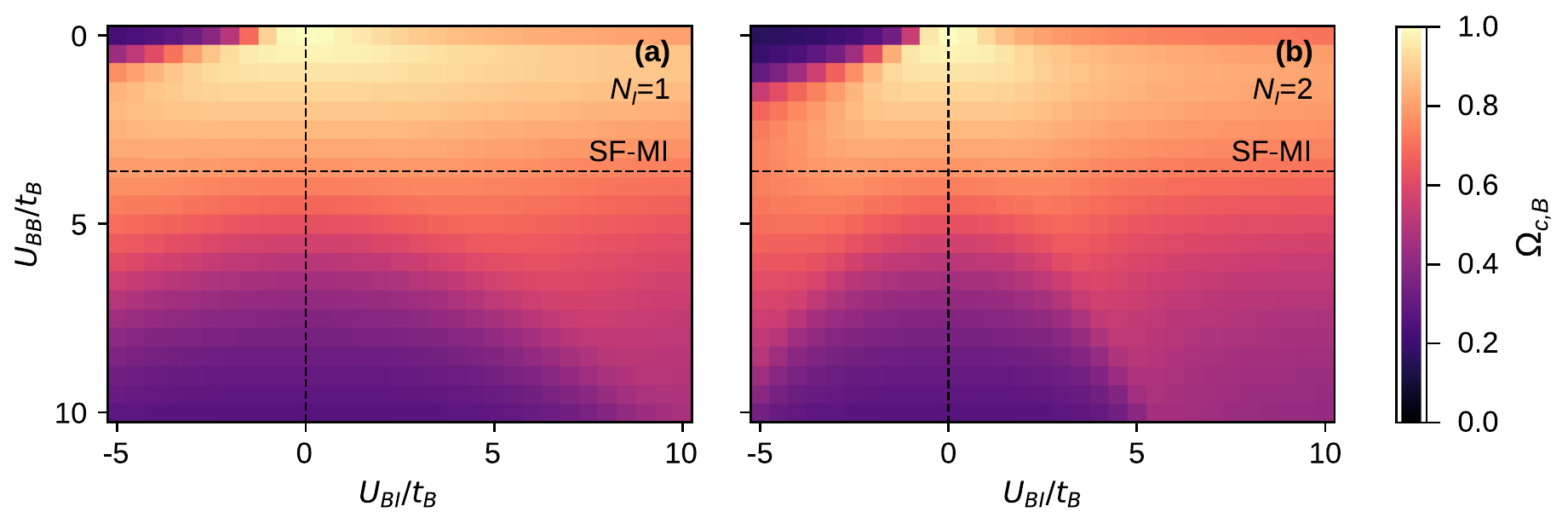}
\caption{Bath's condensate fraction $\Omega_{c,B}$ for $M=9$ and $\nu_B=1$ as a function of $U_{BB}/t_B$ and $U_{BI}/t$. We show results for systems with one impurity (a) and two impurities (b). The horizontal lines indicate the estimated SF-MI phase transition point of the bath $U_{BB}/t_B\approx 3.6$~\cite{kuhner_phases_1998}.}
\label{app:Wc;fig:Ωc}
\end{figure*}

We show condensate fractions of baths interacting with one and two impurities in figure~\ref{app:Wc;fig:Ωc}. As expected, the behavior of the entropy maps onto that of the condensate fraction. Indeed, for vanishing $U_{BI}$, the condensate fraction noticeably decreases for $U_{BB}/t>3.6$, signaling the SF to MI transition~\cite{zhang_exact_2010}. In addition, and similarly to what we reported in the main text, finite boson-impurity interactions with an increasing $|U_{BI}|$ disturb the insulator phase, resulting in smaller regions with $\Omega_{c,B}\approx 0$. Note that, once again, the transition is better defined with two impurities. In addition, the condensate fraction also signals the collapse of the system [upper left corners in both panels], where $\Omega_c$ vanishes, as expected.

%\References
\section*{References}
\bibliography{biblio}

\end{document}